\documentclass[11pt]{article}

\usepackage{amsmath,amssymb}
\usepackage{epsfig}
\usepackage{graphicx}
\usepackage{bm}

\newcommand{\A}{{\bf A}}
\newcommand{\E}{{\bf I}}
\newcommand{\G}{{\bf G}}

\newcommand{\n}{{\bf n}}
\renewcommand{\j}{{\bf j}}
\renewcommand{\k}{{\bm{k}}}
\newcommand{\q}{{\bm{q}}}
\renewcommand{\r}{{\bm{r}}}
\newcommand{\rp}{{\bm{r'}}}
\newcommand{\I}{{\rm i}}

\def\gsim{\lower.35em\hbox{$\stackrel{\textstyle>}{\textstyle\sim}$}}

\begin{document}
\title{Plasmons in layered structures including graphene}
\author{T. Stauber and G. G\'omez-Santos\\\\
{\it Departamento de F\'{\i}sica de la Materia Condensada  and Instituto Nicol\'as Cabrera,}\\{\it Universidad Aut\'onoma de Madrid, E-28049 Madrid, Spain}}
\date{\today}
\maketitle
\abstract{
We investigate the optical properties of layered structures with graphene at the interface for arbitrary linear polarization at finite temperature including full retardation by working in the Weyl gauge. As a special case, we obtain the full response and the related dielectric function of a layered structure with two interfaces. We apply our results to discuss the longitudinal plasmon spectrum of several single and double layer devices such as systems with finite and zero electronic densities. We further show that a nonhomogeneous dielectric background can shift the relative weight of the in-phase and out-of-phase mode and discuss how the plasmonic mode of the upper layer can be tuned into an acoustic mode with specific sound velocity.}\\
\\ 
{Pacs: 78.67.Wj, 73.21.Ac, 42.25.Bs, 73.20.Mf}
\section{Introduction}
Graphene, the two-dimensional allotrope of carbon, has become one of the most active fields in today\rq s both experimental and theoretical condensed matter physics\cite{Geim07,CastroNeto09,DasSarma11,Kotov11}. Among many extraordinary phenomena and proposals, the optical properties of graphene have generated particular interest due to potential applications,\cite{Bonaccorso10,Mueller10,Sun10,Liu11} but also because they are intimately related to the discovery of exfoliated graphene.\cite{Novoselov04}

Recently, plasmonics based on graphene has become a new emerging subfield, trying to take advantage of the strong electronic confinement and long propagation lengths of its carriers and, most importantly, the possibility of applying an electrostatic gate voltage.\cite{Ju11,Koppens11,Moreno11,Seyller08,Tegenkamp11,Basov11} Due to the linear spectrum, different gate voltages can have strong effects on the carrier concentration and thus on the plasmonic spectrum, especially for small nano-islands.\cite{Chen12,Fei12} 

There is also renewed focus on layered structures due to the experimental advances of exfoliating a number of materials and placing them on top of each other.\cite{Ponomarenko11,Britnell12,Kim12} Like this, the Coulomb drag of closely separated graphene layers was observed,\cite{Kim11} which has triggered considerable attention by various theoretical groups.\cite{Katsnelson11,Peres11,Hwang11,Narozhny12,Carrega12,Amorim12,Matos12,Schuett12,Song12} In this paper, we want to combine these two fields and analyze the plasmonic spectrum of layered structures at finite temperature.

To discuss the undamped plasmon dispersion, it suffices to determine the zeros of the dielectric functions\cite{Hwang09} or, alternatively, the zeros of the denominator of the transmission or reflection amplitude.\cite{Stauber12} But for a finite imaginary part, i.e., especially for finite temperatures, this procedure is ambiguous. We will thus need a different approach and will investigate the energy loss function. For single layer graphene systems, the energy loss function is related to the (negative) imaginary part of the inverse of the dielectric function, but for double layer structures, this function changes its sign.\cite{Stauber12} We will thus define the energy loss function as the trace of the (negative) imaginary part of the full response.

Another common approximation to obtain the longitudinal plasmon dispersion consists in using the static Coulomb potential and for small wave numbers $q\ll k_F$ in the regime of small layer separation $k_Fd\ll1$, the Coulomb potential can be further simplified, only depending on the average of the outer dielectric media $(\epsilon_1+\epsilon_3)/2$. For the general case, the full electrostatic problem has to be considered, which has been recently done in the context of graphene double layers.\cite{Stauber12,Profumo12,Badalyan12} 

Since for small energies of the order of $\alpha\epsilon_F$ ($\alpha$ being the fine-structure constant) retardation effects lead to strong light-matter interaction,\cite{GomezSantos12} we will here derive the full retarded photon propagator for the longitudinal and transverse channel. Another focus of this work is placed on materials with a large dielectric constant which strongly screen the graphene layers. SrTiO${}_3$, e.g., has a relative dielectric constant of $\epsilon\sim300$, which can reach up to $\sim5000$ at liquid helium temperature due to the proximity of a ferroelectric instability.\cite{Couto11} Another example are surface states of a three-dimensional topological insulator which are separated by the width of the sample. For Bi${}_{2}$Te${}_3$, the two electronic Dirac systems are then electrostatically coupled through a dielectric medium with $\epsilon\sim100$.\cite{Profumo12} Particularly, we will show that strong dielectrics shift the relative weight of the in-phase and out-of-phase mode. Used as a substrate, they strongly screen the graphene layers and therefore basically act like metals, leading to a linear plasmon dispersion.\cite{Principi11,Politano11} 

The paper is organized as follows. In section II, we derive the photon propagator in free space, separating the final result into longitudinal and transverse channels. In section III, we formulate the linear response theory for a layered structure including graphene in the interfaces and give explicit expressions for the double layer geometry. In section IV, we finally discuss the near-field optical properties of layered graphene structures, contrasting between single and double layer, high- and low temperature, large and small dielectric constants. We close with a summary and an outlook. An appendix outlines the analytical discussion how to obtain the linear plasmon dispersion for a double layer system with large-$\epsilon$ substrate. 

\section{Photon propagator in a homogeneous medium}
\label{Propagator}
In this section and the following section, we will derive the retarded photon propagator in a homogeneous and nonhomogeneous medium, outlining all details. For alternative introductions to confined photon systems, we refer the reader to Ref. \cite{Benisty99}. Even though our treatment will be entirely classical based on Maxwell's equations, we will still call the final result, Eq. (\ref{PhotonPropagator}), the photon propagator since this expression coincides with the quantum mechanical photon propagator.\cite{Landau80} The reader only interested in the results presented in Sec. IV may skip this and the following section.

We depart from Ampere's circuital law equation including Maxwell's displacement current assuming a harmonic time evolution with frequency $\omega$:
\begin{align}\label{Maxwell}
\nabla\times{\bf H}(\r)=-\I\omega{\bf D}(\r)+{\bf j}(\r)\;
\end{align}  
We will first recall the usual representation of the photon propagator in Cartesian coordinates and then introduce the representation in cylindrical coordinates, suitable to discuss layered structures.

\subsection{Photon propagator in Cartesian coordinates}
Introducing the electrostatic potential $\phi$ and the vector potential $\A$ by
\begin{align}
\label{Efield}
{\bf E}(\r)=\I\omega\A(\r)-\nabla\Phi(\r)\\
{\bf H}(\r)=\frac{1}{\mu\mu_0}\nabla\times\A(\r)
\end{align}
we obtain from Eq. (\ref{Maxwell}) the following equation:
\begin{align}
\label{MaxwellAPhi}
\nabla\times\nabla\times\A+\I\omega\mu\mu_0\epsilon\varepsilon_0(\I\omega\A-\nabla\phi)=\mu\mu_0{\bf j}
\end{align}
\subsubsection{Lorentz gauge}
In the Lorentz-gauge $\nabla\cdot\A=\I\omega\mu\mu_0\epsilon\varepsilon_0\phi$, Eq. (\ref{MaxwellAPhi}) can be written as four independent inhomogeneous Helmholtz equations
\begin{align}
(-\nabla^2-k_0^2)A^\mu(\r)=j^\mu(\r)
\end{align}
with the four-dimensional vector potential $A^\mu=(\phi,\A)$ and the four-dimensional current $j^\mu=(\rho/\epsilon\varepsilon_0,\mu\mu_0{\bf j})$. The dispersion relation reads
\begin{align}
k_0^2=\omega^2\mu\mu_0\epsilon\varepsilon_0=\frac{\omega^2}{c^2}\mu\epsilon=\frac{\omega^2}{c_1^2}\;,
\end{align}
where $c$ denotes the speed of light in vacuum and $c_1=c/\sqrt{\mu\epsilon}$ the (slower) speed of light in the dielectric medium characterized by $\mu$ and $\epsilon$. The Green's function defined by 
\begin{align}
(-\nabla^2-k_0^2)G_0(\r,\rp)=\delta(\r-\rp)
\end{align}
is thus a scalar and reads
\begin{align}
G_0(\r,\rp)=\frac{e^{\pm\I k_0|\r-\rp|}}{4\pi|\r-\rp|}\;,
\end{align}
where the plus-sign defines propagation out of the source and the minus-sign convergence into the source. Within the Lorentz gauge, the electric field due to a given current density ${\bf j}$, as defined in Eq. (\ref{Efield}),  is thus given by
\begin{align}
{\bf E}(\r)&=\I\omega\left(1+\frac{\nabla\nabla\cdot}{k_0^2}\right)\A(\r)\label{ElectricFieldLorentzGauge}
=\I\omega\mu\mu_0\left(1+\frac{\nabla\nabla\cdot}{k_0^2}\right)\int d^3r'G_0(\r,\rp){\bf j}(\rp)\;.
\end{align}
\subsubsection{Weyl gauge}
In the following, we will not work in the Lorentz gauge, but will set the electrostatic potential equal to zero, i.e., $\phi=0$. This gauge condition is often refered to as the ``Weyl gauge''. The Weyl gauge implies that we only need to consider the propagation of the vector potential and is often used when interactions with non-relativistic particles are involved. Graphene's linear response to the incoming light field is thus entirely defined by its current density. With $\phi=0$, Eq. (\ref{MaxwellAPhi}) becomes
\begin{align}
\nabla\times\nabla\times\A-k_0^2\A=\mu\mu_0{\bf j}\;.
\end{align}
In this gauge, the operator acting on $\A$ is a vector that connects different spacial directions. For a general solution, we thus need to determine the {\em dyadic} Green's function defined by
\begin{align}
(\nabla\times\nabla\times-k_0^2)\G=\E\delta(\r-\rp)\;,
\end{align}
where $\E$ is the $3\times3$ unit tensor. At first glance, it seems that the Weyl gauge results in more complicated expressions of the Green's function. This is true in real space, but in Fourier space, the expressions are only slightly more involved. And by having eliminated the electrostatic potential $\phi$, the boundary conditions of the subsequent scattering problem become more compact since they only have to be satisfied by the vector potential $\A$.   

We can easily obtain the dyadic Green's function by noting that the Maxwell's equations yield the following expression for the electric field: 
\begin{align}
\nabla\times\nabla\times{\bf E}-k_0^2{\bf E}=\I\omega\mu\mu_0{\bf j}
\end{align}
The electric field is thus defined by the same dyadic Green's function as the vector potential in the Weyl gauge. Since the electric field is gauge independent, we can use the expression of Eq. (\ref{ElectricFieldLorentzGauge}) to deduce the dyadic Green's function $\G$ from the scalar Green's function $G_0$. In real space, we obtain
\begin{align}
\G(\r,\rp)=\left(\E+\frac{\nabla\nabla\cdot}{k_0^2}\right)G_0(\r,\rp)
\end{align}
and in Fourier space
\begin{align}
G^{\alpha,\beta}(\k,\omega)=\left(\delta_{\alpha,\beta}-\frac{k^\alpha k^\beta}{k_0^2}\right)G_0(\k,\omega)\;.
\end{align}
With $G_0(\k,\omega)=(k^2-k_0^2)^{-1}$, we finally obtain the retarded photon propagator $\mathcal{D}_0^{\alpha \beta}=-\mu\mu_0G^{\alpha,\beta}$ in the Weyl gauge,
\begin{align}
\label{PhotonPropagator}
\mathcal{D}_0^{\alpha \beta}(\k,\omega)=\frac{\mu\mu_0}{(\omega/c_1)^2-k^2}\left(\delta_{\alpha,\beta}-\frac{k^\alpha k^\beta}{k_0^2}\right)\;.
\end{align}
Decomposing it in longitudinal and transverse components, it reads
\begin{align}
\mathcal{D}_0^{\alpha \beta}(\k,\omega)=D_L^0\frac{k^\alpha k^\beta}{k^2}+D_T^0\left(\delta_{\alpha,\beta}-\frac{k^\alpha k^\beta}{k^2}\right)
\end{align}
with the functions $D_{L,T}(\k,\omega)$ given by
\begin{align}
D_L^0=\frac{1}{\epsilon\varepsilon_0\omega^2}\;,\;D_T^0=-\frac{\mu\mu_0}{k^2-(\omega/c_1)^2}\;.
\end{align}
\subsection{Photon propagator in cylindrical coordinates}
In a layered structure, assumed to be perpendicular to the $z$ axis, the components parallel to the interface, $\q=(q_x,q_y)$, will be preserved as a good quantum number. It is, therefore, convenient to employ the following representation for the Green's function in a homogeneous medium:
\begin{align}\label{D0zz'}
 \mathcal{D}_0^{\alpha \beta}(z,z';\bm q,\omega) = \frac{1}{2\pi}
 \int dk_z \, {\rm e}^{\I k_z (z-z')} \, 
 \mathcal{D}_0^{\alpha \beta}(\bm k,\omega)
,\end{align} 
with $\bm k = (\bm q,k_z)$.

In this representation, the tensor components have a different structure depending on whether $\alpha,\beta=i,j$ with $i,j=x,y$ or $\alpha,\beta=z$. For the in-plane components of the tensor $\mathcal{D}_0^{\alpha,\beta}$, we have
\begin{align}
\mathcal{D}_0^{i j}(z,z';\q,\omega)=-\frac{\mu\mu_0}{2q'}{\rm e}^{-q'|z-z'|}\, 
		\left(\delta_{i j}-\frac{q_i q_j}{k_0^2}\right)
\end{align}
with $q' = \sqrt{q^2-(\omega/c_1)^2}$. Decomposed into longitudinal and transverse contributions, we obtain
\begin{align}\label{D0explicit}
 \mathcal{D}_0^{i j}(z,z') = d_{l}^0 {\rm e}^{-q'|z-z'|}\,
                                          \frac{q_i q_j}{q^2} + 
					  d_{t}^0 {\rm e}^{-q'|z-z'|}\, 
		\left(\delta_{i j}-\frac{q_i q_j}{q^2}\right)   
,\end{align} 
with the in-plane propagators $d_{l,t}^0(\bm q,\omega)$ given by
\begin{align}\label{dlt}
d_l^0 = \frac{q'}{2 \epsilon\varepsilon_0 \omega^2}\;, \;
d_t^0 = -\frac{\mu\mu_0}{2q'}\;.
\end{align} 
For the cross terms of  $\mathcal{D}_0^{\alpha,\beta}$, we get
\begin{align}\label{D0izexplicit}
 \mathcal{D}_0^{i z}(z,z') = \mathcal{D}_0^{z i}(z,z')=
  d_l^0\frac{\I q_i}{q'}{\rm e}^{-q'|z-z'|} \,{\rm sgn}(z-z')\;.
\end{align} 
This shows that in-plane longitudinal sources can generate not only in-plane, but also out-of-plane fields with a phase shift of $\pi/2$. On the other hand, in-plane transverse currents can only generate in-plane transverse fields.

Finally, out-of-plane sources generate out-of-plane fields given by the tensor
component
\begin{align}\label{D0zzexplicit}
 \mathcal{D}_0^{z z}(z,z') =
 d_l^0\left(2\frac{\delta(z-z')}{q'} - 
   \frac{q^2}{q'^2}{\rm e}^{-q'|z-z'|}\right)\;.
\end{align} 

\section{Linear response of a layered geometry including graphene}
We now consider the experimentally important situation of layered structures, i.e., we assume the existence of well-defined interfaces at which the material properties are discontinuous. The Maxwell equations written in integral form then yield boundary conditions for the normal and tangential field components, see e.g. Ref. \cite{Novotny06}. For the normal components they read
\begin{align} 
\label{BoundaryP}
\n\cdot({\bf D}_2-{\bf D}_1)=\rho\;,\;\n\cdot({\bf B}_2-{\bf B}_1)=0\;
\end{align}
with $\rho$ the charge density on the interface.

For the tangential components they read
\begin{align} 
\label{BoundaryS}
\n\times({\bf E}_2-{\bf E}_1)=0\;,\;\n\times({\bf H}_2-{\bf H}_1)=\j\;
\end{align}
with $\j$ the current density on the interface. 

An arbitrarily polarized electromagnetic wave can always be expressed by superposing two linearly polarized waves which are orthogonal to each other. We can thus define $p$-polarized and $s$-polarized waves, respectively, according to the plane of incidence. Alternatively, we will use the denomination of longitudinal and transverse polarization.

The boundary conditions for the normal and tangential field components are not independent of each other since they are connected by Maxwell's equations. According to the plane of incidence, we will either use Eqs. (\ref{BoundaryP}) in the case of longitudinal polarization or Eqs. (\ref{BoundaryS}) in the case of transverse polarization. These conditions then simplify considerably when working in the Weyl gauge. 

Notice that the influence and properties of graphene are only accounted for via the charge and current densities on the interface, $\rho$ and $\j$. The properties of graphene thus enter when matching the vector field at the interface of two adjacent dielectric media. Since we have set $\phi=0$, these properties are entirely contained in the current-current response of graphene, $\chi_{ij}^0$. The superindex $0$ denotes the bare current response, i.e., the response to the total (external plus induced) vector potential.

We will now assume an isotropic system such that current-current response tensor can be split up into a longitudinal and a transverse contribution, i.e.,
\begin{align}\label{chiij}
\chi_{i j}^0 = \chi_l^0 \frac{q_i q_j}{q^2} + 
             \chi_t^0(\delta_{i j}-\frac{q_i q_j}{q^2})\;.
\end{align} 
Within the Dirac approximation this decomposition is always possible and only for transitions close to the van Hove singularity the full tensor structure needs to be considered.\cite{Stauber10} 

The full photon propagator in the presence of a single graphene layer at the location $z_1$ modifies the ``vacuum'' propagator in the following way:
\begin{align}\label{Dgraphene}
 \mathcal{D}^{\alpha \beta}(z,z') =
 \mathcal{D}_0^{\alpha \beta}(z,z') + 
 \mathcal{D}_0^{\alpha i}(z,z_1)
 \chi_{ij}
 \mathcal{D}_0^{j \beta}(z_1,z')\;,
\end{align} 
where summation over repeated indices is implied. $\chi_{i j}$ represents the total current-current response of graphene to external fields which, decomposed into longitudinal and transverse contributions, is  given by
\begin{align}\label{chifull}
\chi_{i j} = \frac{\chi^{0}_l}{1 -  d_l^0 \chi^{0}_l} \frac{q_i q_j}{q^2} + 
             \frac{\chi^{0}_t}{1 -  d_t^0 \chi^{0}_t} (\delta_{i j}-\frac{q_i q_j}{q^2})\;.
\end{align} 
Longitudinal and transverse components thus decouple and in the following, we will only explicitly label these different channels when it is necessary. The full photon propagator of a general layered structure is obtained by superposing all possible scattering paths similar to Eq. (\ref{Dgraphene}).

Let us finally recall that the gauge field at any point $\r$ due to a current source at $\r'$ is given through the photon propagator via
\begin{align}
  A^\alpha(\r)=-\int d^3r'\mathcal{D}^{\alpha \beta}(\r,\rp)j^\beta(\rp)\;.
\end{align}
Due to this linear correspondence, the photon propagator can be deduced from the scattering problem of the gauge field using the expressions of $\mathcal{D}_0^{\alpha\beta}$ in a homogeneous medium.  

In the following will outline the general scattering problem for one interface, discussing both, the longitudinal and transverse channel. We will then give the explicit expressions for the retarded photon propagator, dielectric function and graphene loss function for a arbitrary double layer structure. 

\subsection{Scattering on one interface}
All general properties of the photon propagator of a layered geometry can be deduced from the scattering problem of one single interface. We will, therefore, discuss this simple problem in some detail. The generalizations are then straightforward. In the following, we will set the plane of incidence the $xz$-plane and the interface at $z=0$. 

\subsubsection{Longitudinal polarization}
For longitudinal polarization, the general vector field has a component parallel ($x$) and normal ($z$) to the interface:
\begin{align}
{\bf A}({\bf r},z)=\sum_\q  e^{\I \q\cdot{\bf r}}\left(A^\parallel(\q,z){\bf e}_\q+A^\perp(\q,z){\bf e}_z\right)\;.
\end{align}
With $q_i^\prime=\sqrt{q^2-(\omega/c_i)^2}$ and $c_i$ the speed of light in the corresponding medium, we make the ansatz $(j=\parallel,\perp)$
\begin{align}
\label{app:AnsatzAlong}
A^j(\q,z)=\left\{
\begin{array}{ll}
a_i^je^{-q_1'z}+a_r^je^{q_1'z}&,z<0\\
a_t^je^{-q_2'z}&,z>0
\end{array}
\right.\;. 
\end{align}
The components of $A^\perp$ are obtained from the components of $A^\parallel$ via the condition for a transverse field $\nabla\cdot{\bf A}=0$. This gives the following relations:
\begin{align}
\label{TransverseRelations}
a_i^\perp=\I\frac{q}{q_1'}a_i^\parallel\;,\;a_r^\perp=-\I\frac{q}{q_1'}a_r^\parallel\;,\;a_t^\perp=\I\frac{q}{q_2'}a_t^\parallel
\end{align}

From Eqs. (\ref{BoundaryP}), we see that the parallel component of the vector field is continuous at the interface, but the normal component of the displacement field makes a jump if a graphene layer is present leading to a charge density $\rho$ at the interface. This component is related to the vector field via the relation $D^\perp=\epsilon\varepsilon_0\I\omega A^\perp$. With the continuity equation $\omega\rho-{\bf q}\cdot{\bf j}=0$ and the linear response $j=-\chi_l^0 A_\q$, the set of equations closes. Together with Eq. (\ref{TransverseRelations}), we thus obtain the following two conditions:
\begin{align}
a_i^\parallel+a_r^\parallel&=a_t^\parallel\\\epsilon_2q_1'a_t^\parallel-\epsilon_1q_2'(a_i^\parallel-a_r^\parallel)&=\frac{q_1'q_2'}{\varepsilon_0\omega^2}\chi_l^0 a_t^\parallel
\end{align}
The transmission and reflection amplitude for longitudinal polarization then read
\begin{align}
T&=\frac{a_t^\parallel}{a_i^\parallel}=\frac{2q_2^\prime\epsilon_1}{q_2^\prime\epsilon_1+q_1^\prime\epsilon_2-\frac{q_1^\prime q_2^\prime\chi_l^0(\q,\omega)}{\varepsilon_0\omega^2}}\;.\\\label{Rlongitudinal}
R&=\frac{a_r^\parallel}{a_i^\parallel}=\frac{q_2^\prime\epsilon_1-q_1^\prime\epsilon_2+\frac{q_1^\prime q_2^\prime\chi_l^0(\q,\omega)}{\varepsilon_0\omega^2}}{q_2^\prime\epsilon_1+q_1^\prime\epsilon_2-\frac{q_1^\prime q_2^\prime\chi_l^0(\q,\omega)}{\varepsilon_0\omega^2}}\;.
\end{align}
\subsubsection{Transverse polarization}
For transverse polarized light and the plane of incidence again in the $xz$-plane, only the $y$-component $A_y$ of the vector field is non-zero. We thus make the following ansatz:
\begin{align}
\label{AnsatzA}
A_y(\r,z)=\sum_\q e^{\I \q\cdot\r}\left\{
\begin{array}{ll}
a_ie^{-q_1'z}+r_ie^{q_1'z}&,z<0\\
t_ie^{-q_2'z}&,z>0
\end{array}
\right. 
\end{align}
From Eqs. (\ref{BoundaryS}) we see that the vector potential is continuous at the interface and that the first derivative makes a jump due to the current generated by the vector field inside the graphene plane. The current is again related to the corresponding transverse current-current susceptibility, $\chi_t^0$, via linear response.\cite{Principi09,Stauber10} We thus obtain the following two conditions:
\begin{align}
a_i+a_r&=a_t\\
-\frac{q_2'}{\mu_2}a_t-\frac{q_1'}{\mu_1}(a_r-a_i)&=\mu_0\chi_t^0 a_t
\end{align}
The transmission and reflection amplitude for transverse polarization then read
\begin{align}
T=\frac{a_t}{a_i}&=\frac{2\mu_2q_1^\prime}{\mu_2q_1^\prime+\mu_1q_2^\prime+\mu_1\mu_2\mu_0\chi_t^0(\q,\omega)}\;,\\
R=\frac{a_t}{a_i}&=\frac{\mu_2q_1^\prime-\mu_1q_2^\prime-\mu_1\mu_2\mu_0\chi_t^0(\q,\omega)}{\mu_2q_1^\prime+\mu_1q_2^\prime+\mu_1\mu_2\mu_0\chi_t^0(\q,\omega)}\;.
\end{align}

\begin{figure}[t]
\begin{center}
  \includegraphics[angle=0,width=0.5\linewidth]{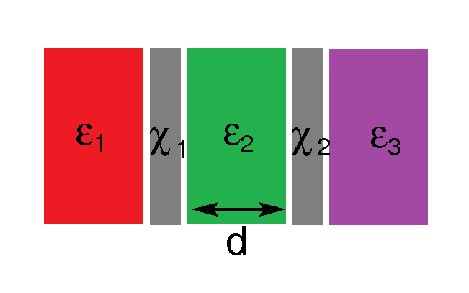}
\caption{(color online): Schematic setup of the double layer graphene structure. The two graphene layers, characterized by graphene's current response of the two layers $\chi_1$ and $\chi_2$ and separated by a distance $d$, are sandwiched by the three dielectric media characterized by $\epsilon_1$, $\epsilon_2$, and $\epsilon_3$. The corresponding magnetic permeabilities $\mu_1$, $\mu_2$, and $\mu_3$ are suppressed.} 
  \label{figure1}
\end{center}
\end{figure}
\subsection{Scattering on two interfaces}
We will now discuss in some detail the special case of a double layer graphene structure by applying the basic matching conditions discussed in the previous subsection to two interfaces. We will give explicit expressions for the retarded photon propagator, the dielectric function and the full response of graphene including the definition of the energy loss function.
 
\subsubsection{Photon propagator}
For two and more interfaces it is convenient to write Fourier transformed part of the photon propagator as a $n\times n$-matrix, $n$ denoting the number of interfaces. The full photon propagator then satisfies the usual Dyson-like equations (one for each polarization channel) 
\begin{equation}\label{RPAphoton} 
\bm d = (\bm 1 -  \bm d^0 \bm \chi^0)^{-1} \bm
d^0  ,\end{equation}   
where for the special case of two graphene layers the matrix $\bm \chi^0 =
\text{diag}(\chi_1^0,\chi_2^0)$ represents the bare  graphene's
response in layer 1 ($\chi_1^0$) and layer 2 ($\chi_2^0$). $\bm d^0$ is the photon propagator in the absence of graphene ($\chi_i^0=0$), but with the dielectric geometry of Fig. \ref{figure1}.

The entries of the matrix $\bm d$ can be obtained from the standard
 matching conditions or, equivalently, using multiple scattering formalism.
 In the latter case, they can be written as
\begin{align}\label{entries} 
 (\bm d)_{11}&=d_1\,(1+\tilde{r}_{1,3}) &(\bm d)_{12}&=d_3\,\tilde{t}_{3,1} \\
 (\bm d)_{21}&=d_1\,\tilde{t}_{1,3}     &(\bm d)_{22}&=d_3\,(1+\tilde{r}_{3,1})
,\end{align} 
where $d_i=\tfrac{q'_i}{2 \epsilon_i \varepsilon_0 \omega^2}$ for longitudinal polarization, and $d_i=-\tfrac{\mu_i \mu_0}{2 q'_i}$ for the transverse case, see Eq. (\ref{dlt}). The compound reflection and transmission amplitudes are 
\begin{equation}
\label{MultipleInterfaces}
\begin{split}
\tilde r_{1,3} & = r_{12} + \frac{t_{12} r_{23} t_{21} 
\text{e}^{-2 q'_2 d}}{1 -r_{21} r_{23} \text{e}^{-2 q'_2 d} } \\
\tilde t_{1,3} &= \frac{t_{12} t_{23}\text{e}^{ -q'_2 d}}{1 -r_{21} r_{23} \text{e}^{-2 q'_2 d}}
\end{split}
,\end{equation}  
with the obvious expression for index exchange,  
where $r_{ij} (t_{ij})$ are the corresponding coefficients for a single
graphene interface from medium $i$ into medium $j$, previously obtained.
Notice that a recursive  interpretation of the last formula can be used for
calculating the reflection and transmission amplitudes for any layered structure
with multiple interfaces.     

Solving the matching conditions for the scattering problem with two interfaces but without graphene ($\chi^0=0$) directly, one can obtain more explicit expressions for the photon propagator in the absence of graphene, ${\bf d}^0$.
Using the results of Ref. \cite{Stauber12}, one obtains for longitudinal polarization the following compact result:
\begin{align}
{\bf d}_l^0=\frac{q_1'q_2'q_3'\epsilon_2}{\varepsilon_0\omega^2N_{l,0}}
\left(
\begin{array}{cc}
\cosh(q_2'd)+\frac{q_2'\epsilon_3}{q_3'\epsilon_2}\sinh(q_2'd)& 1\\
1& \cosh(q_2'd)+\frac{q_2'\epsilon_1}{q_1'\epsilon_2}\sinh(q_2'd)
\end{array}
\right)
\end{align}
with $N_{l,0}=q_2'\epsilon_2(q_3'\epsilon_1+q_1'\epsilon_3)\cosh(q_2'd)+({q_2'}^2\epsilon_1\epsilon_3+q_1'q_3'\epsilon_2^2)\sinh(q_2'd)$.
For the transverse part, one obtains the corresponding expression
\begin{align}
{\bf d}_t^0=-\frac{\mu_1\mu_2\mu_3\mu_0q_2'}{N_{t,0}}
\left(
\begin{array}{cc}
\cosh(q_2'd)+\frac{\mu_2q_3'}{\mu_3q_2'}\sinh(q_2'd)& 1\\
1& \cosh(q_2'd)+\frac{\mu_2q_1'}{\mu_1q_2'}\sinh(q_2'd)
\end{array}
\right)
\end{align}
with $N_{t,0}=\mu_2q_2'(\mu_3q_1'+\mu_1q_3')\cosh(q_2'd)+(\mu_2^2q_1'q_3'+\mu_1\mu_3{q_2'}^2)\sinh(q_2'd)$. 
\subsubsection{Graphene's response}
Graphene's response obeys  similar equations (one for each polarization channel)
\begin{equation}\label{RPAchi}
\bm \chi = (\bm 1 -  \bm \chi^0 \bm d^0)^{-1} \bm \chi^0 
,\end{equation}  
which, together with Eq. (\ref{RPAphoton}), provide the complete dynamics of 
the coupled matter-field system. The retarded dielectric function of double layer graphene with nonhomogeneous background is then often defined by\cite{Hwang09}
\begin{align}
\label{Dielectric}
\epsilon(q,\omega)=\det(\bm 1 - \bm \chi^0 \bm d^0)\;. 
\end{align}
Usually $-\text{Im}\epsilon^{-1}$ is used to discuss the plasmonic spectrum, but this function changes sign and can thus not be interpreted as (positive definite) spectral density. But instead of the determinant, we find it more convenient to discuss the trace of the the full response matrix. Graphene's excitations correspond to the imaginary part of the full response, and to reveal its presence we will discuss the following generalization of the energy loss function $S(q,\omega)$ to several layers:
\begin{equation}\label{loss}
S(q,\omega)=-\text{Im}\chi(q,\omega) = -\frac{1}{e^2}\text{Im Tr}\bm \chi(q,\omega)\;
\end{equation}  

Since $S(q,\omega)$ is related to the imaginary part of a causal function, it is strictly positive and since it also proportional to the usual definition of the energy loss function for a single layer, Eq. (50) can serve as a straightforward generalization of the energy loss function for arbitrary layered systems. 

Let us briefly comment on the physical interpretation of the response matrix and the related energy loss function. The diagonal entries of the response matrix $\bm\chi$ are given by the response of a particular layer if the gauge field is only applied to just this particular layer. Diagonalizing the response matrix $\bm\chi$, one can discuss the elementary excitations of the full system, separately. This was done in Ref. \cite{GomezSantos12}, where for double-layer graphene the in-phase and out-of-phase excitations were analyzed. Since the trace of the response matrix $\bm\chi$ is invariant with respect to unitary transformations, it serves as natural choice for the definition of the generalized energy loss function and for the discussion of the internal excitations of the whole system, i.e., the sum of excitations of all layers. 

As mentioned above, the imaginary part of the general response is related to graphene's excitations. In the following section, we will use the developed formalism to discuss the spectrum of longitudinal plasmonic excitations for single and double layer structures. For this, retardation can formally be neglected. But we stress that the formalism can also straightforwardly be used for multiple layer structures as well as to discuss the spectrum of transverse plasmonic excitations where retardation effects are crucial.
\begin{figure}[t]
\begin{center}
  \includegraphics[angle=0,width=0.8\linewidth]{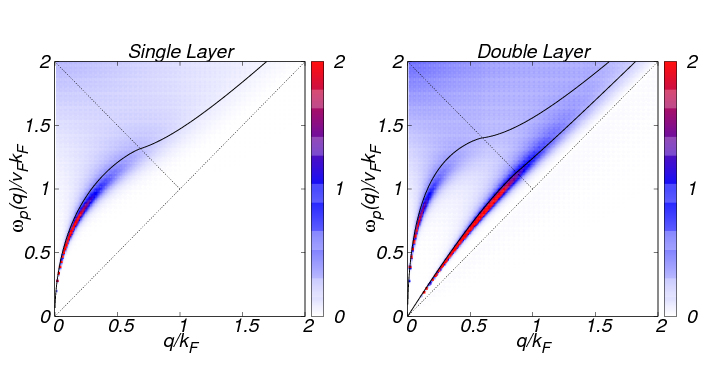}
  \caption{(color online): The energy loss function $-\text{Im}\chi(q,\omega+\I0)$ in units of $\epsilon_F/\hbar^2$ for longitudinal polarization for single layer (left) and double layer with $k_Fd=0.35$ (right) at a temperature $T=T_F/4$ with the same dielectric medium for all regions $\epsilon=1$ (air). Also shown the zero-temperature plasmon dispersion (black solid lines).} 
  \label{figure2}
\end{center}
\end{figure}

\begin{figure}[t]
\begin{center}
  \includegraphics[angle=0,width=0.8\linewidth]{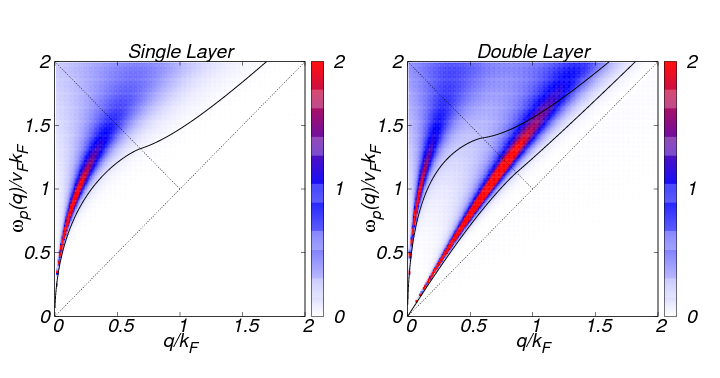}
  \caption{(color online): The energy loss function $-\text{Im}\chi(q,\omega+\I0)$ in units of $\epsilon_F/\hbar^2$ for longitudinal polarization for single layer (left) and double layer with $k_Fd=0.35$ (right) at a temperature $T=T_F$ with the same dielectric medium for all regions $\epsilon=1$ (air). Also shown the zero-temperature plasmon dispersion (black solid lines).} 
  \label{figure3}
\end{center}
\end{figure}
\begin{figure}[t]
\begin{center}
  \includegraphics[angle=0,width=0.8\linewidth]{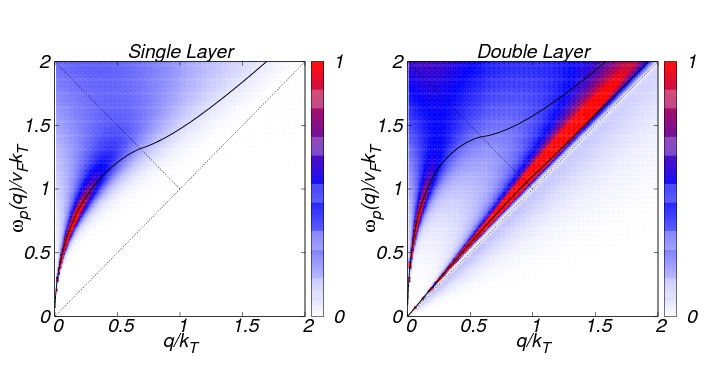}
  \caption{(color online):  The energy loss function $-\text{Im}\chi(q,\omega+\I0)$ in units of $\epsilon_T/\hbar^2$ ($\epsilon_T=\hbar v_Fk_T$) for longitudinal polarization for single layer (left) and double layer with $d=2$nm (right) at a temperature $T=300$K at zero doping $n=0$ with the same dielectric medium for all regions $\epsilon=1$ (air). Also shown the zero-temperature plasmon dispersion for finite doping with $k_T=\tfrac{k_BT}{\hbar v_F}2\ln2$ (black solid lines).} 
  \label{figure4}
\end{center}
\end{figure}
\section{Plasmons in layered structures at finite temperature}

\begin{figure}[t]
\begin{center}
  \includegraphics[angle=0,width=0.8\linewidth]{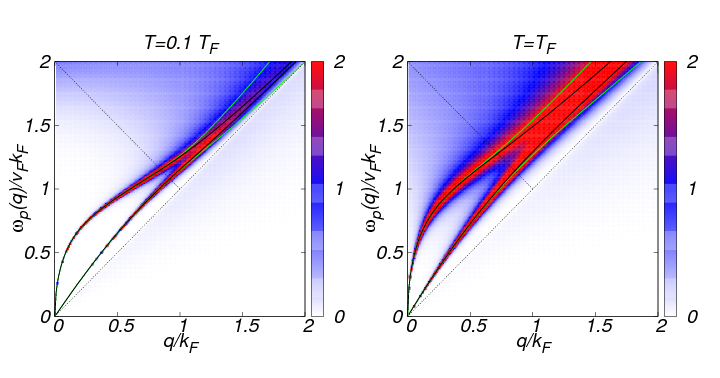}
  \caption{(color online): The energy loss function $-\text{Im}\chi(q,\omega+\I0)$ in units of $\epsilon_F/\hbar^2$ for longitudinal polarization of double layer graphene with equal carrier density and layer separation $k_Fd=1.77$ at temperature $T=T_F/10$ (left) and $T=T_F$ (right) with $\epsilon_1=1$, $\epsilon_2=6$ and $\epsilon_3=3.8$. Also shown the finite-temperature plasmon dispersion obtained by Re(det$\epsilon)=0$ for Im$P^0=0$ (black lines) and finite Im$P^0$ (green lines).} 
  \label{figure5}
\end{center}
\end{figure}
\begin{figure}[t]
\begin{center}
  \includegraphics[angle=0,width=0.8\linewidth]{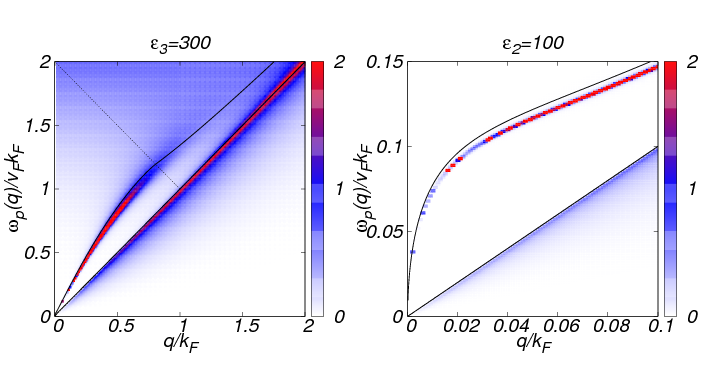}
  \caption{(color online): The energy loss function $-\text{Im}\chi(q,\omega+\I0)$ in units of $\epsilon_F/\hbar^2$ and temperature $T=T_F/4$ for longitudinal polarization. Right hand side: Double layer graphene with $k_Fd=0.35$ with large dielectric substrate $\epsilon_3=300$ ($\epsilon_1=\epsilon_2=1$). Left hand side: topological insulator ($g_v=g_s=1$) with width $k_Fd=2.37$ and $\epsilon_2=100$ ($\epsilon_1=1$, $\epsilon_3=4$). Also shown the zero-temperature plasmon dispersion (black solid lines).} 
  \label{figure6}
\end{center}
\end{figure}
In this section, we will apply our formalism and discuss the longitudinal response of layered structures at finite temperature. To do so, we will use the density-density correlation function $P^0(q,\omega)$ which is related to the longitudinal component of the current-current correlation function $\chi_l^0(q,\omega)$ via the continuity equation.\cite{Stauber10} Within the Dirac-cone approximation, this reads
\begin{align}
  \chi_l^0(q,\omega)=\frac{\omega^2}{q^2}P^0(q,\omega)\;.
\end{align}

\subsection{Polarizability of graphene}
Up to now, there is no approximation involved except of decomposing the current response into a longitudinal and transverse contribution which is well justified for any transitions not too close to the van Hove singularity. We will now approximate the full density-density correlation function by the non-interacting polarizability\cite{Wunsch06}
\begin{align}
P^{0}(\mathbf{q},i\omega_n)&=\frac{g_s g_v}{4\pi^2}\int
d^2k\sum_{s,s^{\prime}=\pm}f^{ss^{\prime}}(\k,\mathbf{q})\,
\frac{n_F(E^{s}(k))-n_F(E^{s^{\prime}}(|\mathbf{k} + \mathbf{q}|))}{%
E^{s}(k)-E^{s^{\prime}}(|\mathbf{k}+\mathbf{q}|)-i\hbar \omega_n}\,,
\label{eq:P1}
\end{align}
with $E^{\pm}(k)=\pm \hbar v_F k$ the eigenenergies, $n_F(E)=(e^{\beta
(E-\mu)}+1)^{-1}$ the Fermi function, and $g_s=g_v=2$ the spin and valley
degeneracy for graphene. A characteristic difference between the polarizability of graphene and that of a two-dimensional electron gas is the appearance of the prefactors $f^{ss^{\prime}}(\k,\mathbf{q})$ coming from the band-overlap of the wave function
\begin{align}
f^{s s^{\prime}}(\k,\mathbf{q})&=\frac{1}{2}\left(1+s s^{\prime}\frac{%
k+q\cos\varphi}{|\k+\mathbf{q}|}\right),
\end{align}
where $\varphi$ denotes the angle between $\k$ and $\mathbf{q}$.

At zero temperature, we have $\mu=\epsilon_F$ with $\epsilon_F$ the Fermi energy and the analytic solution of Ref. \cite{Wunsch06} can be decomposed in several patches corresponding to inter- and intraband transitions, respectively. In Figs. 2-6, the three most basic patches are separated by black dashed lines.

At finite temperature the chemical potential is determined with respect to the electronic density $n$ by the following relation:
\begin{align}
\label{muT}
\int_{-\infty}^{\infty} d\epsilon\nu(\epsilon)\left[n_F(\epsilon)-\Theta(-\epsilon)\right]=n\;,
\end{align}
with the density-of-states given by  $\nu(\epsilon)=g_sg_v|\epsilon|/(2\pi v_F^2)$. For an electron-doped system, we have $\epsilon_F>\mu>0$. At the neutrality point, $n=0$ and due to particle-hole symmetry we then have $\mu=0$. For our numerical calculations, we will make use of the semi-analytical expression of the polarizability presented in Ref. \cite{Ramezanali09}.

\subsection{Loss function at finite and zero doping}
We will now consider single and double layer structures at finite and zero  doping at low and high temperatures. To clarify the discussion we will choose a homogeneous medium, i.e., $\epsilon_1=\epsilon_2=\epsilon_3=1$.

We first discuss the plasmon dispersion at finite temperature for suspended single and double layer graphene with a layer separation of $k_Fd=0.35$. For an electronic density of $n=10^{12}$cm${}^{-2}$, we then have $d=2$nm and $T_F=1200$K, such that $T=T_F/4$ would correspond to approximately room temperature. The energy loss function is shown for $T=T_F/4$ in Fig. \ref{figure2} and for $T=T_F$ for \ref{figure3}. 

We compare the energy loss function with the plasmon dispersion at zero temperature by determining det$\epsilon=0$. In the case of a finite imaginary part of $P^0(q,\omega)$, i.e., Landau damping, we set Im$P^0$=0. This guarantees the convergence of the two plasmonic modes for large $q$.\cite{Stauber12} We see that at intermediate temperatures the plasmon dispersion is red-shifted compared to the zero temperature solution, whereas for high temperatures it is blue-shifted. This follows directly from the behavior of $P^0(q=0,\omega\to0)$ which is related to the Drude weight $D$ and defines the plasmon dispersion.\cite{Wunsch06} At finite temperature, this reads
\begin{align}
D=\frac{g_sg_v}{\pi}\frac{k_BT}{\hbar^2}\left(\ln(1+e^{\mu/k_BT})+\ln(1+e^{-\mu/k_BT})\right) 
\end{align}
which is a non-monotonic function of the temperature due to the temperature dependence of the chemical potential given in Eq. (\ref{muT}).

In Fig. \ref{figure4}, we show the plasmon dispersion at the neutrality point $n=0$ at $T=300$K. It was already discussed earlier that also in this limit (damped) plasmon excitations exist in single layer graphene due to the weak Landau damping.\cite{Vafek06} In fact, there is an analytical approximation which uses the zero temperature result of the polarizability\cite{Wunsch06} with the replacement $k_F\rightarrow k_T=\tfrac{k_BT}{\hbar v_F}2\ln2$.\cite{Falkovsky07} This agrees with the physical expectation that at finite temperature there is a finite electronic density due to thermal broadening of the Fermi function. 

The parameter in Fig. \ref{figure4} corresponds to a thermal wave number $k_T=0.061$nm${}^{-1}$ and electron density $n=1.2\times10^{11}$cm$^{-2}$. The energy loss function is given in units of $\epsilon_T/\hbar^2$ with $\epsilon_T=\hbar v_F k_T\sim36$meV. As one can see, the analytical approximation agrees well with the maximum of $-\text{Im}\chi(q,\omega+\I0)$ even in the case of double layer graphene.

\subsection{Loss function for nonhomogeneous dielectric media}
We will now discuss the effect of a nonhomogeneous dielectric media which can lead to changes of the plasmon dispersion due to the different photon propagator. In Fig. \ref{figure5} we show the results for a double layer graphene structure for two temperatures $T=T_F/10$ (left) and $T=T_F$ (right). We choose SiO$_2$ with $\epsilon_3=3.8$ as a substrate and Al$_2$O$_3$ with $\epsilon_2=6$ as a buffer layer. It is further assumed that the top layer is air with $\epsilon_1=1$. The two layers have the same electron density and layer separation $k_Fd=1.77$ which for $n=10^{12}$cm$^{-2}$ corresponds to $d=10$nm.

We use the same parameters as in Ref. \cite{Badalyan12} and the results of this reference are shown as green solid lines, obtained from the real part of det$\epsilon=0$. Since the authors do not set the imaginary part to zero, the in-phase and out-of-phase modes do not merge for $q\gsim k_F$ in contrary to the physical expectation. The correct way to approximate the plasmon dispersion from the zeros of the dielectric function is thus to set the imaginary part of the current-current correlation function to zero (black solid lines). Even after including temperature broadening and Landau damping, the discussion of Ref. \cite{Badalyan12} remains incorrect since it is based on the response of single layer graphene.

Apart from the mere (numerical) corrections of the plasmonic dispersion due to a nonhomogeneous dielectric background, it is possible to shift spectral weight from the out-of-phase to the in-phase mode. One can also change the plasmon dispersion from the typical square-root dispersion of charged plasmonic waves to a linear dispersion typical for acoustic sound waves. In Figs. \ref{figure2}-\ref{figure4}, where the same dielectric constant was chosen to be the one of air, the out-of-phase mode is more dominant extending towards larger wave numbers. In Fig. \ref{figure5}, we see that the two modes are almost equal in weight merging in the region of finite Landau damping. If we choose now a substrate with a very large dielectric constant, the in-phase mode becomes more dominant whereas the out-of-phase mode basically merges with the Dirac cone. Since the latter behavior is also seen in the bare graphene response, we can thus state that there is only an in-phase plasmonic mode. This is shown on the left hand side of Fig. \ref{figure6}.

The plasmonic mode is also dominant in the case of a topological insulator where the buffer layer is resembled by a strong dielectric medium, i.e., we set $\epsilon_1=1$, $\epsilon_2=100$, and $\epsilon_3=4$. This can be seen on the right hand side of Fig. \ref{figure6} where we use the same parameters as in Ref. \cite{Profumo12}. They are given by $g_v=g_s=1$ with a sample width of $k_Fd=2.37$ and a Fermi velocity of $v_{TI}=5\times10^5$m/s which scales out.

Let us finally comment on the linear plasmon mode present on the left hand side of Fig. \ref{figure6}. In the appendix, we give details of how to derive and approximate the plasmon dispersion of the upper layer as $\omega_p=v_aq$. For this we do not include the full expression of the current-current correlation function, but only use the long wavelength limit. This yields the simple result for the group velocity of the linear (acoustic) mode
\begin{align}
\label{groupvelocity}
v_a=\frac{2\alpha_gdk_F^1}{\epsilon_2}v_F\;,
\end{align}
where $k_F^1$ is the Fermi wave number of the upper graphene layer and $\alpha_g=\alpha c/v_F$ graphene's fine-structure constant. This approximation breaks down for small $dk_F$ and large $\epsilon_2$ since $v_a$ cannot become smaller than the Fermi velocity\cite{Stauber12} and a more careful analysis is necessary.\cite{Profumo12} 

Eq. (\ref{groupvelocity}) can nevertheless be used to discuss several aspects. First, $v_a$ only depends on the buffer substrate; second, only the upper graphene layer enters in the expression since the lower one is perfectly screened. This formula can thus not only be applied to double layer graphene on substrates with large dielectric constants, but also to single layer graphene on top of metals interpreting $d$ as the graphene-metal distance. We can thus calculate the group velocity of the plasmonic mode for a single layer graphene on top of Pt(111) as discussed experimentally in Ref. \cite{Politano11} where the energy loss function was measured. With the parameters $\epsilon_F=0.3$eV, $d=0.33$nm and $\epsilon_2=1$ corresponding to the experimental set-up, we have $k_Fd=0.15$ and thus $v_a=1.15v_F$. The screened plasmon dispersion will thus lie close to the Dirac ``light-cone'' in agreement with experiment.

\section{Summary}
We presented results of the energy loss function of layered structures which is the only way to unambiguously discuss the plasmon dispersion in the presence of dissipative terms like Landau damping or finite temperature. We discussed the effect of temperature and nonhomogeneous dielectric medium, including large dielectrics which almost perfectly screen the graphene layers, but our formalism equally applies to systems with unbalanced electronic densities. Our main results are (i) the possibility of shifting relative weight of the several plasmonic branches by changing the nonhomogeneous dielectric background and (ii) a simple formula for the sound velocity of the (linear) plasmonic mode of the upper graphene layer in the case of a substrate with large dielectric constant. These insights might be useful towards the engineering of specific plasmon modes for future plasmonic circuitries based on graphene.

\section{Acknowledgments}
This work has been supported by Portugal's Funda\c{c}\~ao para a Ci\^encias e a Tecnologia (FCT) via grant PTDC/FIS/101434/2008, by Spain's Ministerio de Ciencias e Innovaci\'on (MICINN) via grant FIS2010-21883-C02-02 and by Spain's Ministerio de Economia y Competitividad (MINECO) via grant FIS2012-37549-C05-03.

\section{Appendix: Large substrate screening and acoustic plasmons}

In this appendix, we show that a large (huge) value of $\epsilon_3$ renders medium 3 almost a  metal, largely screening the Coulomb interaction in the upper graphene layer and turning the otherwise square-root plasmon into an acoustic mode over a wide wave number range. This is similar to the proposal of Ref. \cite{Principi09} where a perfect metal as substrate is considered. The acoustic mode evolves into a regular two-dimensional plasmon only at much reduced wave numbers, due to the incomplete screening of medium 3.

\subsection{Linear plasmon mode of the upper graphene layer}
We assume the usual geometry where the three dielectric media separate the upper and lower graphene layer, see Fig. \ref{figure1}. We assume a very large value of $\epsilon_3 \approx 300$, much greater than the other regular values of $\epsilon_{1,2}$, and ignore retardation. Plasmons are solutions of (see Eq. (\ref{MultipleInterfaces}))
%
\begin{align}
1  - r_{21} r_{23}\text{e}^{- 2 q d}  = 0 
.\end{align}  
The reflection amplitude $r_{23}$ is given by (see Eq. (\ref{Rlongitudinal}) in the limit $c\rightarrow\infty$)
\begin{align}\label{eqrij} 
r_{23}  = 
\frac{\epsilon_2  - \epsilon_3  +  \frac{q \chi_2^0}{ \varepsilon_0\omega^2}} 
{\epsilon_2 
+ \epsilon_3  - \frac{q \chi_2^0}{ \varepsilon_0\omega^2}} 
.\end{align} 
In the limit $\epsilon_3 \gg \epsilon_{1,2}$, $r_{23}\approx -1$, i.e., there is perfect screening of the lower graphene layer. On the other hand $r_{21}$,
given by
\begin{align} 
r_{21}  = 
\frac{\epsilon_2  - \epsilon_1  +  \frac{q \chi_2^0}{ \varepsilon_0\omega^2}} 
{\epsilon_2 
+ \epsilon_1  -  \frac{q \chi_2^0}{ \varepsilon_0\omega^2}} 
,\end{align}
can be rewritten upon ignoring the $q$ dependence of $\chi_1^0\rightarrow\frac{e^2}{\hbar}\frac{g_sg_v}{4\pi}v_Fk_F$ (local response) as 
\begin{align} 
r_{21}  = \frac{
\frac{\epsilon_2-\epsilon_1}{\epsilon_2+\epsilon_1}
\omega^2 + \omega^2_{p}}{\omega^2 - \omega^2_{p}}
,\end{align} 
where $\omega^2_{p}$ is the plasmon dispersion for the upper graphene layer between the semi-infinite dielectrics $\epsilon_{1,2} $, given by   
\begin{align}
\omega^2_{p} =  \frac{1}{(\epsilon_1+\epsilon_2)\varepsilon_0} \chi_1^0 q
.\end{align} 
Now, the solution of $1  + r_{21}\text{e}^{- 2 q d}  = 0$ in the limit 
$ q \rightarrow 0$ is
\begin{align}
\omega^2 \rightarrow \frac{\epsilon_1+\epsilon_2}{2\epsilon_2}  \omega^2_{p} 2
q d
.\end{align} 
This is clearly an acoustic mode. The physics is simple: medium 3 acts just as a
metal, screening the long-range Coulomb interactions in upper graphene layer,
turning the original square-root plasmon into an acoustic mode. 

\subsection{Range of validity: Regular plasmon}

For sufficiently small $q$, the screening cannot be perfect and the long range nature of the interaction should show up anyway. The previous analysis uses the approximation $r_{23}=-1 $. One can see that the first correction to lowest order in $\frac{\epsilon_{1,2}}{\epsilon_3} $ is 
\begin{align}
r_{23}(q=0, \omega\rightarrow 0) = -1 + \frac{2 \epsilon_2}{\epsilon_3}
,\end{align} 
and using this result, the range of validity of the acoustic regime can be
established as
\begin{align}
\frac{\epsilon_2}{\epsilon_3} << q d \lesssim 1
.\end{align} 
For smaller wave numbers $ q d \lesssim \frac{\epsilon_2}{\epsilon_3} $, one
recovers the standard square root behavior due to long-range Coulomb
interactions, albeit with much reduced frequencies. To the alluded order, one
easily finds the {\em regular} plasmon as
\begin{align}
\omega^2 \approx \frac{\epsilon_1+\epsilon_2}{\epsilon_3}\omega^2_{p},\;\;
\text{for}\; \;q d \lesssim \frac{\epsilon_2}{\epsilon_3}
.\end{align} 

\end{document}